\documentclass[12pt]{iopart}
\usepackage{iopams}
\newcommand{\sump}{\mathop{{\sum}'}_{n=0}^\infty }
\begin{document}
\title[Surface and photonic modes in Casimir calculations]
{Surface modes and photonic modes in Casimir calculations  for a compact cylinder}

\author{V V Nesterenko}

\address{Bogoliubov Laboratory of Theoretical Physics, Joint Institute for Nuclear Research,
Dubna 141 980, Russia}
% \address{$^2$ Department of Mathematics,
% Imperial College, Prince Consort Road, London SW7~2BZ, UK}
% \address{$^3$ Department of Computer Science,
%University College London, Gower Street, London WC1E~6BT, UK}
\ead{nestr@theor.jinr.ru}

\begin{abstract}
A rigorous formulation of the problem  of calculating  the
electromagnetic vacuum energy of an infinite dielectric cylinder
is discussed. It is shown that the physically relevant  spectrum
of electromagnetic excitations includes the surface modes and
photonic modes. The mathematical procedure of summing over this
spectrum is proposed, and the transition to imaginary
frequencies is accomplished. As a result, it is justified the
imaginary-frequency representation for the vacuum energy which
has been used in previous Casimir studies for this
configuration.
\end{abstract}
% \submitto{\JPA}
%\maketitle
\section{Introduction}
The notion of the elementary excitation spectrum is of paramount
importance in all condensed matter  physics~\cite{LL,Kittel}.
The excitations of different type result, as a rule, in
different physical consequences. Therefore it is of a certain
interest for theoretical and experimental investigations of the
Casimir effect to answer  the question: the electromagnetic
oscillations of what type are considered in the problem at
hand~\cite{Ford-1,Ford-2,Henkel,Genet,Lambrecht,Bordag}. Having
elucidated this point one can hope to link, in a transparent
way, the Casimir force with actual physical properties of the
material boundaries. However it is not easy to answer this
question even when the Casimir force is calculated by making use
of the familiar Lifshitz formula~\cite{Lif,DLP1,DLP2}. A rather
complicated derivation of this formula~\cite{ST1} in the
original papers initiated its obtaining anew  by making use
mainly of the mode
summation method~\cite{ST1,ST2,PL1,NPW,PRB,PL2,Lang,KMM}.

The Casimir calculations for nonflat boundaries turned out to be
much more involved in comparison with those for
planes~\cite{RNC}. Especially complicated calculations have been
done for a circular cylinder~\cite{DeRaad-Milton,BN,DeRaad,MNN,LNB,NP1,NP,BP,%
ICP-Milton1,ICP-Milton2,ICP-Milton3,Barton,GRomeo,Klich-Romeo,%
Romeo-Milton,Romeo-Milton-Barcelona,Brevik-Romeo,Schaden}.
It was also unclear electromagnetic excitations of what kind have been
taken into account in these studies~\cite{Barcelona}.

The present work seeks to present a consistent derivation of
the formula for the vacuum energy of electromagnetic field
connected with a material cylinder by summing explicitly the
contributions to this energy given by  different branches of the
electromagnetic spectrum in this problem.

The layout of the paper is as  follows. In Sec.\ 2 the spectral
problem generated by the Maxwell equations for a compact
infinite cylinder is formulated  rigorously and the physically
relevant spectrum of electromagnetic excitations for this
configuration is determined. It is shown that this spectrum
includes surface modes (bound states) and photonic modes. In
Sec.\ 3 the summation over this spectrum is accomplished by
making use of the spectral density when accounting of the
photonic (continues) branch  of the spectrum. In Conclusion
(Sec.\ 4) the meaning of obtained results are discussed  briefly.

\section{Physical spectrum of electromagnetic excitations for a
cylinder}
In the source-free case the general solution to Maxwell equations
can be
represented in terms of two independent Hertz vectors~\cite{Stratton}
\begin{eqnarray}
\mathbf{E}&=&\bnabla\times \bnabla\times {\bPi}'
+i\,\mu\,\frac{\omega}{c}\, \bnabla \times \bPi''\,{,}
\label{2-5}\\
\mathbf{H}&=&-i\,\varepsilon
\,\frac{\omega}{c}\,\bnabla\times
{\bPi}'+\bnabla \times {\bnabla}\times {\bPi}''\,{.}
\label{2-6}
\end{eqnarray}
Here $\mathbf{\Pi '}$ is the electric Hertz vector, $\mathbf{\Pi
''}$ is  the  magnetic Hertz vector, $c$ is the velocity of
light in vacuum, and the Gauss units are used. The Hertz vectors
obey the Helmholtz vector equation
\begin{equation}
\left (  {\bnabla}^2 + k^2\right ){\bPi}=0\,{,} \label{2-7}
\end{equation}
where the wave number $k$ is given by
\begin{equation}
\label{2-8} k^2=\varepsilon \mu\,\frac{\omega^2}{c^2}\,{.}
\end{equation}

On the other hand, it is  known that  the general solution to
Maxwell without sources equations can be derived from two scalar functions which
may be chosen in deferent ways~\cite{Whittaker,Nisbet}. For the
configuration with cylindrical symmetry the role of these
functions can play the axial components of  the electric
($\mathbf{\Pi '}$)  and  magnetic ($\mathbf{\Pi ''}$) Hertz
vectors. The rest components of $\mathbf{\Pi '}$ and
$\mathbf{\Pi ''}$ are zero in this case. As a result, the
Helmholtz vector equation (\ref{2-7}) reduces to the scalar
Helmholtz equations for  $ {\bPi}'_z\equiv {\Pi}'$ and
${\bPi}''_z\equiv {\Pi}''$
\begin{equation}
\label{2-9} \left ( \Delta +\varepsilon \mu \frac{\omega^2}{c^2}
\right ) \Pi =0, \qquad \Pi=\Pi',\;\Pi''
\end{equation}
with the following general solutions
\begin{eqnarray}
\Pi '=\sum_{n=0,\pm1,\pm2,\ldots}
a_nf_n^{\rm{TM}}(r)e^{ihz+in\theta},
\label{2-10} \\
\Pi ''=\sum_{n=0,\pm1,\pm2,\ldots} b_n
f_n^{\rm{TE}}(r)e^{ihz+in\theta}. \label{2-11}
\end{eqnarray}
The cylindrical coordinates ($r,\theta,z$) are used and  the $z$
axis coincides  with the axis of a circular infinite cylinder of
radius $a$. The medium inside the cylinder has the permittivity
$\varepsilon_1$ and permeability $\mu_1$. These quantities
outside the cylinder acquire the values $\varepsilon_2$ and
$\mu_2$, respectively. We assume for definiteness that
$\varepsilon_1\mu_1> \varepsilon_2\mu_2$. The wave vector along the  $z$ axis is
denoted by $h$.
The amplitudes $a_n$ and $b_n$ for the solutions inside the cylinder will be denoted by
$a_n^i$ and $b_n^i$, respectively, and in the same way for solutions outside the cylinder
we introduce the amplitudes $a_n^e$ and $b_n^e$.

The functions $f^{\rm {TE}}_n(r)$ and $f^{\rm {TM}}_n(r)$ in
the general solutions (\ref{2-10}) and (\ref{2-11}) obey the
radial wave equation
\begin{equation}
\fl \label{2-12} \frac{d^2
f_n}{dr^2}+\frac{1}{r}\frac{df_n}{dr}+\left (
k^2-h^2-\frac{n^2}{r^2} \right)f_n=0,\qquad
f_n(r)=f^{\rm{TE}}_n(r),\,f^{\rm {TM}}_n(r)\,{.}
\end{equation}
Inside the cylinder we put
\begin{equation}
\label{2-13} f_n(r)=J_n(\lambda_1 r), \qquad n=0,1,\ldots,\qquad
0<r<a\,{,}
\end{equation}
where $J_n(z)$ is the Bessel function, $\lambda_1 =\sqrt{k_1^2-h^2},\,
k_1^2=\omega^2/c_1^2=\varepsilon_1\mu_1\omega^2/c^2$.
Outside the cylinder we consider first  'outgoing' waves
\begin{equation}
 \label{2-15} f_n(r)=H^{(1)}_n(\lambda_2 r),\qquad n=0,1,\dots, \qquad r>a\,{,}
\end{equation}
where $H^{(1)}(z)$ is the Hankel function of the first kind, $\;
\lambda_2=\sqrt{k_2^2-h^2},\;
k_2^2=\omega^2/c_2^2=\varepsilon_2\mu_2\,\omega^2/c^2$.

In the radial solutions (\ref{2-13}) and  (\ref{2-15}) the sign of $\lambda_s^2=k_s^2-h^2,\;s=1,2$
is not fixed yet. Thus, in our consideration
the solutions
\begin{equation}
\label{ad-1}
f_n(r)=I_n(\bar \lambda_1r) \qquad \mbox{for} \qquad r<a
\end{equation}
and
\begin{equation}
\label{ad-2}
f_n(r)=K_n(\bar \lambda_2r)   \qquad \mbox{for} \qquad  r>a
\end{equation}
are also admissible. Here $\bar \lambda _s^2=h^2-k^2_s, \quad
s=1,2$,  $I_n(z)=\rmi^{-n} J_n(\rmi z)$ and $K_n(z)=\rmi^{n+1}\frac{\pi}{2}H_n(\rmi z)$ are the
modified Bessel functions~\cite{GR}.

On the cylinder surface
the matching conditions should be satisfied. These conditions require the
continuity of tangential components of fields $\mathbf{E}$ and
$\mathbf{H}$ when crossing cylinder surface
\begin{equation}
\label{6-4} \mbox{discont } (\mathbf{E}_{\parallel}) =0,\qquad
\mbox{discont } (\mathbf{H}_{\parallel})=0 \,{.}
\end{equation}

The matching conditions give rise to
the frequency  equation determining admissible values of the spectral parameter
$\omega$ in the boundary value problem under consideration:
\begin{eqnarray}
\fl \label{2-21} \frac{\omega^2a^4}{c^2}\left ( \varepsilon_1
\lambda _2\frac{J_n'}{J_n} -\varepsilon_2 \lambda
_1\frac{H_n'}{H_n}
 \right )
\left ( \mu_1 \lambda _2\frac{J_n'}{J_n} -\mu_2 \lambda
_1\frac{H_n'}{H_n}
 \right ) \nonumber \\
 - \frac{n^2h^2a^2}{\lambda_1^2 \lambda_2^2} \left [
\frac{\omega^2}{c^2} (\varepsilon_1\mu_1-\varepsilon_2\mu_2)
\right ]^2 =0\,{,}\quad n=0,1,2, \ldots \,{.}
\end{eqnarray}
In this equation
\begin{equation}
\fl
 J_n\equiv
J_n(\lambda_1 a),\qquad H_n\equiv H_n^{(1)}(\lambda_2 a), \qquad
 \lambda_s=+\sqrt{
{\omega^2}/{c_s^2}-h^2},\qquad s=1,2\,{,}
\label{2-22}
\end{equation}
the prime on the  functions $J_n$ and $H_n$ means
differentiation with respect to their arguments, $c_1$ and $c_2$
are the velocities of light inside and outside the cylinder,
respectively, $c_2>c_1$.

The roots of  (\ref{2-21}) are important in
radio-engineering when developing the radio dielectric
waveguides~\cite{Rayleigh,Hondros,Hondros1,HondrosDebye,Sch,Kac,HdP1}
and in fiber optics (optical waveguides~\cite{Snitzer,Marcuse}).
The results of investigation of the  frequency equation
(\ref{2-21}) determining the spectrum in the problem at hand
can be summarized in the following way. All the
{\it real roots} of this equation  lie
in the interval
\begin{equation}
\label{2-24} c_1h<\omega < c_2 h\,{.}
\end{equation}
These roots make up two discrete sequences. In the interval (\ref{2-24}) the frequency
equation (\ref{2-21}) can be rewritten
in the form
\begin{eqnarray}
\fl  \frac{\omega^2a^4}{c^2}\left ( \varepsilon_1 \bar
\lambda _2\frac{J_n'}{J_n} +\varepsilon_2 \lambda
_1\frac{K_n'}{K_n}
 \right )
\left ( \mu_1 \bar \lambda _2\frac{J_n'}{J_n} +\mu_2 \lambda
_1\frac{K_n'}{K_n}
 \right )\nonumber \\-
\frac{n^2h^2a^2}{\lambda_1^2 {\bar \lambda_2}^2} \left [
\frac{\omega^2}{c^2} (\varepsilon_1\mu_1-\varepsilon_2\mu_2)
\right ]^2 =0, \qquad n=0,1,2,\ldots \label{2-25}
\end{eqnarray}
with the notation $
K_n\equiv K_n(\bar \lambda_2 a)$.

When  frequency $\omega$ equals the real roots of  equation
(\ref{2-25}), located in the interval (\ref{2-24}), the 'outgoing' waves (\ref{2-15})
become the functions decaying in the radial direction  (\ref{ad-2}) (surface or
evanescent waves).
These eigenmodes describe the propagation of  electromagnetic
waves along  the cylinder (waveguide solutions).
The radial functions (\ref{ad-1}) are not realized in the problem under consideration.

Now we address the complex roots of the frequency equation
(\ref{2-21}). In the strip of the complex frequency plane
\begin{equation}
\label{2-39} 0< {\rm Re}\;\omega < c_2 h
\end{equation}
there are no complex roots of (\ref{2-21})  with ${\rm Im}\; \omega
\neq 0$.

In the semi-plane
\begin{equation}
\label{2-33} {\rm Re}\,\omega  > c_2  h
\end{equation}
for sure there are {\it complex roots} of (\ref{2-21}) with
${\rm Im}\; \omega \neq 0$. Indeed, in the domain (\ref{2-33})
the left-hand side of (\ref{2-21}) is a complex function of the
complex variable~$\omega$. The complex eigenfrequencies of a
dielectric cylinder  lead to leaky (radiating) modes. It is
clear that these modes cannot carry the electromagnetic
energy along the cylinder. For us it is important that the modes
with complex $\omega$ (quasi-normal modes~\cite{Barcelona}) do not satisfy
standard completeness condition and as a result they cannot be
used for quantization of electromagnetic field  in the
problem at hand.

In order to get rid of the complex eigenfrequencies and
consequently to escape  leaky or radiating modes we shall
consider,  outside the cylinder, the scattering states instead
of outgoing waves. The scattering solutions to Maxwell equations can be derived
from outgoing solutions by
making use of the substitutions
\begin{equation}
\eqalign{
a_n^e H_n^{(1)}(\lambda_2 r)&\to a_n^+ H_n^{+}(\lambda_2 r)+
a_n^- H_n^{-}(\lambda_2 r)  \cr
b_n^e H_n^{(1)}(\lambda_2 r) &\to b_n^+ H_n^{+}(\lambda_2 r)+
b_n^- H_n^{-}(\lambda_2 r)\cr}\,{.} \label{2-19}
\end{equation}
For simplicity in (\ref{2-19})
the notations
\begin{equation}
\label{2-20} H_n^{+}\equiv H_n^{(1)}, \quad H_n^{-}\equiv
H_n^{(2)}
\end{equation}
are introduced.

 As a result, for a given $n$ and $h$ we have
6 amplitudes $a_n^i,b_n^i,a_n^{+},a_n^{-},b_n^{+},b_n^{-}$. The
matching conditions at the cylinder surface lead to 4 linear
homogenous equations for these amplitudes. Hence no restrictions
arise here for the spectral parameter $\omega^2/c^2$.

Eliminating in these equations the amplitudes $a_n^i$ and $b_n^i$ we are left
with 2 equations in 4 amplitudes $a_n^{\pm}$ and $b_n^{\pm}$
\begin{equation}
K^-\left(
\begin{array}{c} a_n^+ \\ b_n^+
\end{array}
\right )
=K^+
\left(
\begin{array}{c} a_n^- \\ b_n^-
\end{array}
\right ), \quad K^{\pm}=\mp\left (
\begin{array}{cc}
\alpha^\mp & \beta^\mp \\
\gamma^\mp& \alpha^\mp
\end{array}
\right ){,}
\end{equation}
where
\begin{eqnarray}
\alpha^{\pm}_n=\frac{nh}{a}\left(
-1+\frac{\lambda_2^2}{\lambda_1^2}
\right),\nonumber \\ \beta ^\pm_n= -\rmi \frac{\omega}{c}\left (
\mu_2\lambda_2 {H_n^{\pm}}'-\mu_1\lambda_1\frac{J_n'}{J_n}\,\frac{\lambda_2^2}{\lambda_1^2}
H_n^{\pm}\right ), \nonumber \\
\gamma ^\pm_n= \rmi \frac{\omega}{c}\left (
\varepsilon_2\lambda_2 {H_n^{\pm}}'-\varepsilon_1\lambda_1\frac{J_n'}{J_n}\,\frac{\lambda_2^2}{\lambda_1^2}
H_n^{\pm}\right )\,{.}
\end{eqnarray}

The $S$ matrix in this problem
\begin{equation}
\left(
\begin{array}{c} a_n^+ \\ b_n^+
\end{array}
\right )
=S
\left(
\begin{array}{c} a_n^- \\ b_n^-
\end{array}
\right )
\end{equation}
obeys obviously the following matrix equation
\begin{equation}
K^- S=K^+\,{,}
\end{equation}
and
\begin{equation}
\label{ad-6}
\det S= \frac{\det K^+}{\det K^-}\,{.}
\end{equation}
By a direct calculation one can easily show that $\det K^{-}$
coincides (up to unimportant multiplier) with the left-hand
side of the frequency equation (\ref{2-21}). Thus this equation can be
rewritten in the form
\begin{equation}
\label{ad-7}
\det K^{-}=0\,{.}
\end{equation}
Surprisingly  formulas (\ref{ad-6}) and (\ref{ad-7}) for the $S$
matrix were not  known in the literature devoted to the
electromagnetic scattering by a cylinder.

Summarizing we infer that the spectrum of electromagnetic
oscillations in the problem under study consists of discrete
values $\omega_{n\alpha}$: $c_1 h<\omega_{n\alpha}<c_2h$
corresponding to the surface modes and a continuous branch of
the spectrum with real positive $\omega$: $c_2 h<\omega <
\infty$. In mathematical scattering theory~\cite{Newton} it
is proved that the bound states and scattering states form
together a complete set.

\section{Summation over the spectrum and transition to imaginary
frequencies}
Now we address the calculation of the vacuum energy in the problem at hand
proceeding from the standard mode-by-mode summation
\begin{equation}
\label{3-1}
\fl E_c= \frac{1}{2}\sum_{\{q\}}\omega_q=
\int_{-\infty}^{\infty}\frac{\rmd h}{2\pi}
\sump\left [ \sum_\alpha \omega _{n\alpha}(h) +
\int_{c_2 h}^\infty \omega\,
\Delta \rho_n (\omega,h)\,\rmd \omega
\right ]{,}
\end{equation}
where the prime over the sum sign means that the term with $n=0$
is taken with the weight $\case{1}{2}$. The first term in square
brackets is responsible for the surface waves contribution and the second one describes
the contribution of the photonic modes. The latter contribution
is represented by making use of the respective spectral shift function $\Delta
\rho$~\cite{Barcelona}.
The rigorous mathematical scattering
theory gives the following expression for the spectral density
shift
\begin{equation}
\label{3-2} \Delta \rho (k)\equiv \rho (k)-\rho_0(k)
=\frac{1}{2\pi i}\,\frac{d}{dk}\, \tr \ln
S(k)=\frac{1}{2\pi i}\frac{d}{dk}  \ln \det S(k)\,{.}
\end{equation}
Here $\rho(k)$ is the density of states for a given potential
(or for a given boundary conditions in the case of compound
media) and $\rho_0(k)$ is the spectral density in the respective
free spectral problem (for vanishing potential or for
homogeneous unbounded space). It is obvious that in the Casimir
calculations one has to use just   $\Delta \rho (k)$ subtracting
at this point the so-called Minkowski space-time  contribution
to the vacuum energy.

In the case of scalar scattering problem the Jost functions
$f(k)$ and $f(-k)$,   the scattering matrix $S(k)$,  and the
phase shift $\delta (k)$ are related by the formula
\begin{equation}
\label{3-3} S(k)=e^{2i\delta(k)}=\frac{f(k)}{f(-k)}\,{.}
\end{equation}
Substitution of  (\ref{3-3}) into (\ref{3-1}) gives more
familiar formula for spectral density~\cite{LLStat1}
\begin{equation}
\label{3-4} \Delta \rho (k) =\frac{1}{2\pi i}\,\frac{d}{dk}\,\ln
\frac{f(k)}{f(-k)}=\frac{1}{\pi}\,\frac{d}{dk}\,\delta(k)\,{.}
\end{equation}

In the problem under consideration the TE and TM modes do not decouple.
Therefore we are dealing here with the matrix $(2\times 2)$ scattering problem
and we must use the spectral density defined by (\ref{3-2}).

The contribution of the surface modes in (\ref{3-1}) can be represented by the counter intergral
\begin{equation}
\label{3-5}
\sum_\alpha \omega _{n\alpha}=\frac{1}{2\pi \rmi}\oint_{C}\omega \frac{\rmd}{\rmd \omega}
\ln F_n(\omega) \,\rmd \omega\,{,}
\end{equation}
where $F_n(\omega)$ is the left hand side of (\ref{2-25}). This
equation was written for real $\omega$. However, in the counter
integral (\ref{3-5}) an analytical continuation of this function
to the complex frequency plane should be used. It can be done
immediately in terms of $\det K^+$ (lower semi-plane $\omega$)
and $\det K^-$ (upper semi-plane $\omega$). After that we can
use for both terms in (\ref{3-1}) the counter integral
representations with the counters $C_+$ and $C_-$, respectively.
The counter $C_-$ starts at $\rmi \infty$ and goes along the
positive imaginary axis to the origin and after that it goes
along the positive real semi-axis to infinity. The counter $C_+$
is obtained by refection of $C_-$ to the upper semi-plane
$\omega$. As a result we arrive at the following imaginary
frequency representation of the vacuum energy in the problem at
hand
\begin{equation}
\label{3-6}
 E_c=
\int_{-\infty}^{\infty}\frac{\rmd h}{2\pi}
\sump
\int _0^\infty y \frac{\rmd}{\rmd y}\ln F_n(\rmi y,h)  \rmd y\,{,}
\end{equation}
where $F_n(\omega, h )$ is the left hand side of the frequency
equation (\ref{2-21}). It is this representation that has been
used in the Casimir calculations for a material cylinder.

\section{Conclusion}
We have shown that in the case of a material cylinder there are
two types of electromagnetic excitations which are physically
relevant: i) surface modes and ii) photonic modes. A consistent
transition to imaginary frequencies requires the  both branches
of the spectrum are to be taken into account. The contribution
to the Casimir energy due to the surface modes and photonic
modes can be separated only in terms of real frequencies. Upon
transition to imaginary frequencies these contributions are
indivisible. Presented consideration justifies rigorously the
imaginary frequency representation for  the Casimir energy of a
compact infinite cylinder that has been used in many previous
papers dealing with investigation of this energy.

It is worth noting that the mathematical consideration presented
here is completely applicable to the Lifshitz configuration,
namely, to an infinite dielectric plate placed in vacuum
(dielectric films).

 \ack This study has been accomplished by the  financial support of Russian Foundation for
Basic Research (Grant 06-01-00120) and the  Heisenberg-Landau Program.

\end{document}